\documentclass[twocolumn,showpacs,preprintnumbers,amsmath,amssymb,footinbib]{revtex4-1}

\usepackage{graphicx}
\usepackage{epstopdf,mathrsfs}
\usepackage{booktabs}

\begin{document}


\title{On the relationship between pump chirp and single-photon chirp in spontaneous
parametric downconversion}
\author{Xochitl Sanchez-Lozano$^{1}$, Alfred B. U'Ren$^{2}$, Jose Luis Lucio$^1$}
\affiliation{
$^{1}$Divisi\'on de Ciencias e Ingenier\'ias,
Universidad de Guanajuato, P.O. Box E-143, 37150, Le\'on Gto.,
M\'exico. \\
$^{2}$Instituto de Ciencias Nucleares, Universidad Nacional Aut\'onoma de M\'exico, Apdo. Postal 70-543, M\'exico 04510 DF}

\date{\today}

\begin{abstract}
We study the chronocyclic character, i.e. the joint temporal and spectral properties, of the single-photon constituents of photon pairs generated by spontaneous parametric down conversion. In particular we study how single photon properties, including purity and single-photon chirp, depend on photon pair properties, including the type of signal-idler spectral and correlations and the level of pump chirp.
\end{abstract}

\maketitle

\section{INTRODUCTION}
\label{introduction}

Single photons constitute the most fundamental buil\-ding block of optical fields. Single photons are often described in terms of a single optical mode described by an annihilation operator $\hat{a}$, i.e. as $\hat{a}^\dag | 0 \rangle$, where $|0 \rangle$ is the vacuum. However, single photons emitted through actual physical processes are of course wavepackets involving different optical frequencies as well as different directions of propagation. A complete understanding of single photons emitted under realistic conditions therefore requires an in-depth study of their multi-modal richness.

In this paper we concentrate our study on single photons derived from photon pairs emitted through the process of spontaneous parametric down conversion (SPDC) in second-order non-linear crystals.  Likewise, we focus on the spectral degree of freedom in the SPDC photon pairs. The process of SPDC can be extremely versatile as the basis for photon-pair source design; indeed, a careful selection of the source configuration leads to the ability to widely select the properties of the emitted photon pairs. Thus, for example, photon pair sources exhibiting entanglement in discrete photonic degrees of freedom (such as polarization~\cite{kwiat95,kwiat99} and orbital angular momentum~\cite{mair01}), and alternatively exhibiting entanglement in continuous degrees of freedom~\cite{osorio08,vicent10} (such as frequency/time~\cite{grice97}, transverse momentum/position~\cite{howell04}) have been demonstrated. In the specific case of the spectral degree of freedom, photon-pair sources have been designed and demonstrated with a large range of behaviors in terms of the emission bandwidth (ranging from tens of MHz~\cite{haase09} to hundreds of THz~\cite{odonnell07}) and in terms of the degree of the entanglement (ranging from factorable~\cite{mosley08} to highly entangled~\cite{odonnell07}).

The detection of one of the two photons in a given SPDC pair, can herald the presence of its conjugate, an approach often used as the basis for single-photon sources~\cite{uren04,pittman05}.  Heralded single photons have been characterized in previous works, for example through a measurement of the corresponding Wigner function in the phase space formed by the position and momentum electric field quadratures~\cite{lvovsky01, aichele02}.   In this paper we are primarily interested in studying how the spatial and temporal properties of the pump translate into specific properties of the constituent singles photons, in particular in the time-frequency degree of freedom.   Indeed, in a recent work from our group, we have shown that pump chirp may be used in order to control effectively the degree of entanglement in SPDC photon pairs, and consequently control the purity of the constituent single photons~\cite{jeronimo09}. We have shown that if the source is designed so that in the absence of pump chirp photon pairs are factorable, or equivalently so that the constituent single photons are pure~\cite{uren05}, an arbitrary degree of photon-pair entanglement, or single-photon purity, can be attained by varying the level of pump chirp.

Besides the single-photon purity, in this paper we are also interested in the chronocyclic character, i.e. in the joint spectral-temporal properties, of the single photons.  Specifically, we are interested in studying how the chronocyclic character of the pulse train associated with a broadband pump defines the chronocyclic character of the emitted single photons.    While this is interesting from a fundamental-physics perspective, it is also important in the context  of quantum information processing (QIP) applications which rely on single photons.
Knowledge of the full multi-mode structure of single photons is essential for the correct design of specific QIP implementations.    Indeed, an understanding of, and the ability to control these chronocyclic properties are crucial for the correct mode-matching of single photons to other photonic modes in interferometric arrangements.  For example,
both Hong-Ou-Mandel interference of single photons from distinct sources, and homodyning of single photons with a coherent state depend crucially on the single-photon chronocyclic character. Therefore, it is important to understand how pump chirp, used as a tool for tailoring the level of spectral entanglement present in SPDC photon pairs, also determines the chronocyclic character of the single photons which constitute each pair.

An ideal framework in which to study the single-photon joint spectral-temporal properties is the single-photon Chronocyclic Wigner function (CWF).  In this paper, we study the relationships between the single photon spectral density matrix, the first order degree of spectral and temporal coherence, and the single-photon CWF. We study in particular how the type of spectral correlations in the SPDC photon pairs, together with the level of pump chirp, determine on the one hand the purity and on the other hand the presence of spectral-temporal co\-rre\-la\-tions, or chirp, in the single photons.

\section{SPDC Photon pairs, and their constituent single photons}

In the spontaneous parametric downconversion process (SPDC) an optical crystal with a second-order non-linearity is illuminated by a laser pump beam.  Individual photons from the pump beam may then be annihilated, leading to the emission of photon pairs, where the two photons in a given pair are typically referred to as signal and idler.  These photon pairs may be entangled in any of the photonic degrees of freedom, in\-clu\-ding polarization, time-frequency and transverse position-momentum.   In this paper we concentrate on the spectral degree of freedom; specifically,  we assume that appropriate spatial filtering on the signal and idler modes is used
so that only specific directions of propagation are retained.

Following a standard perturbative approach, the two-photon state for the SPDC process can be written as $|\Psi \rangle =|0\rangle+\eta | \Psi_2 \rangle$, in terms of the two-photon component of the state $| \Psi_2 \rangle$

\begin{align}
\label{1} |\Psi_2\rangle = \int d\omega_s \int d\omega_i f(\omega_s, \omega_i)|\omega_s\rangle|\omega_i\rangle,
\end{align}

\noindent where $\eta$ is  related to the conversion efficiency, $f(\omega_s, \omega_i)$ is the joint spectral amplitude (JSA) function, $|\omega_\mu\rangle = \hat{a}^\dag_\mu(\omega)|0\rangle$
with $\mu=i,s$, and $|0\rangle$ represents the vacuum state.

The focus of this paper is on the properties of the single photons which constitute each of the SPDC pairs.  Specifically, we are interested in studying how the single-photon properties are determined by the photon-pair properties.  Each of the single photons in a given pair is completely characterized by its respective density operator.    For the specific case of the signal mode, the density operator $\hat{\rho}_s$ can be written as

\begin{align}
\label{2} \hat{\rho}_s = \mbox{Tr}_i(|\Psi \rangle \langle \Psi |),
\end{align}

\noindent where $\mbox{Tr}_i$ represents a partial trace over the idler mode. In terms of the JSA,
the reduced density operator can be expressed through its matrix elements as

\begin{align}
\label{3} \rho_s(\omega_1,\omega_2) &\equiv  \langle \omega_1 | \hat{\rho}_s | \omega_2 \rangle \\ \nonumber
&=\int d\omega_0 f(\omega_1,\omega_0)f^*(\omega_2, \omega_0).
\end{align}

The density matrix $\rho_s(\omega_1,\omega_2)$ fully characterizes the signal-mode single photons.  Note that the normalization condition on the density operator $\mbox{Tr}(\hat{\rho}_s)=1$ is equivalent to the normalization condition $\int d \omega_s \int d \omega_i |f(\omega_s,\omega_i)|^2=1$ for the JSA function.  For what follows, it is convenient to write down the density matrix expressed in terms of diagonal $\omega$ and off-diagonal $\omega'$ frequency components, $\rho^D(\omega,\omega')$, where  $\omega$ and $\omega'$ are related to $\omega_1$ and $\omega_2$ through $\omega_1=\omega+\omega'/2$ and $\omega_2=\omega-\omega'/2$,

\begin{equation}
\label{4} \rho_s^D(\omega,\omega')=\rho_s \left(\omega+\frac{\omega'}{2},\omega-\frac{\omega'}{2}\right).
\end{equation}

An important property of the single photons is their purity, quantified by

\begin{align}
\label{5} p \equiv \mbox{Tr}(\hat{\rho}^2)
=\int d \omega_1 \int d \omega_2 |\rho_s(\omega_1,\omega_2)|^2.
\end{align}

Aside from the purity, in this paper we are interested in the spectral and temporal properties of the single photons.  The Wigner distribution function applied to the time-varying electric field provides a mathematical tool which can represent classical light pulses in chronocyclic, i.e. in time-frequency space \cite{paye92}. This leads to a representation that
shows intertwined temporal and spectral effects, aiding a better comprehension of certain optical phenomena.   In the realm of quantum optics, it is convenient to
study the temporal and spectral properties of single photons through the single-photon
Chronocyclic Wigner function~\cite{uren07}. Indeed, while the density matrix contains complete information about the single photons, in this paper we also use the single-photon chronocyclic function (CWF)  $W(\omega,t)$, because this leads to a better direct appreciation of the temporal, as well as spectral, properties of the single photons. In particular, phase information which is absent from the absolute value of the density matrix $|\rho_s(\omega_1,\omega_2)|$, appears naturally in the real-valued CWF.  Certain spectral-temporal properties such as single-photon chirp become more directly apparent in the chronocyclic domain.

The density matrix $\rho(\omega_1,\omega_2)$ and the chronocyclic Wigner function $W(\omega,t)$ are related to each other through a Fourier transform, as follows

\begin{align}\label{6}
W(\omega,t)= \frac{1}{2 \pi}\int d \omega'  \rho_s^D \left(\omega,\omega' \right) e^{-i \omega' t}.
\end{align}

Or, conversely,

\begin{align}
\label{7} \rho_s^D(\omega,\omega')=\int dt \,\, W(\omega,t)e^{i\omega' t}.
\end{align}

It is known that integration of the CWF over the time variable yields the single-photon spectral intensity, or single-photon spectrum (SPS) $I_\omega(\omega)$.  Thus, we can show that the SPS is closely related to the diagonal elements of the density matrix, i.e.

\begin{align}
\label{8} I_\omega(\omega) \equiv \int d t W(\omega,t)= \rho_s(\omega,\omega)= \rho_s^D(\omega,0).
\end{align}

The function $\rho_s(\omega,\omega)$ is then related to the ``population of each single-photon spectral component'', in other words to the relative spectral intensity at each of the spectral components.  Normalization of $I_\omega(\omega)$, so that $\int d \omega I_\omega(\omega)=1$ is guaranteed by the normalization of the density matrix, i.e. $\mbox{Tr}(\rho_s)=1$.  The off-diagonal elements, or ``coherences between different spectral components'', which occur for $\omega' \neq 0$,  then can have non-zero values for pure states and vanish for highly impure states.  In fact, the interpretation of the off-diagonal elements as coherences may be made direct by observing that there is a simple relationship between the first degree of spectral coherence $S(\omega_1,\omega_2)$ between two frequencies $\omega_1$ and $\omega_2$ and the density matrix $\rho(\omega_1,\omega_2)$.  $S(\omega_1,\omega_2)$ may be defined as follows ~\cite{torrescompany09}

\begin{equation}
\label{9} S(\omega_1,\omega_2) \equiv \mbox{Tr}\left[\hat{\rho}_s a^\dag(\omega_1)a(\omega_2)\right].
\end{equation}

It is then a simple matter to show that

\begin{equation}
\label{10} S(\omega_1,\omega_2)=\rho^*(\omega_1,\omega_2).
\end{equation}

Alternatively, we may choose to study the coherence between two different times.  In this case, we define the first degree of coherence $\Gamma(t_1,t_2)$ between two different times $t_1$ and $t_2$.   In terms of the time-domain annihilation operator $\tilde{a}(t)=\int d \omega a(\omega) e^{-i \omega t}$, $\Gamma(t_1,t_2)$ may be defined as

\begin{equation}
\label{11} \Gamma(t_1,t_2) \equiv \mbox{Tr}\left[\hat{\rho}_s \tilde{a}^\dag(t_1)\tilde{a}(t_2)\right].
\end{equation}

It can then be shown that $\Gamma(t_1,t_2)$ may be expressed in terms of the density matrix $\rho^D(\omega,\omega')$ as

\begin{equation}
\label{12} \Gamma(t_1,t_2) = \int d \omega \int d \omega' \rho^D(\omega,\omega') e^{-i\frac{\omega'}{2}(t_1+t_2)}e^{i \omega (t_1-t_2)},
\end{equation}

\noindent or alternatively  we may also relate $\Gamma(t_1,t_2)$  to the CWF $W(\omega,t)$  through

\begin{equation} \label{13}
\Gamma\left(t+\frac{t'}{2},t-\frac{t'}{2}\right) = 2\pi \int d \omega W(\omega,t) e^{i \omega t'},
\end{equation}

\noindent which corresponds to a form of the Wiener-Khintchine theorem for single photons \cite{torrescompany09}.

Thus, the Fourier transform of $\rho^D(\omega,\omega')$ with respect to the anti-diagonal frequency component $\omega'$ yields the Wigner function $W(\omega,t)$, while the Fourier transform of the Wigner function with respect to the diagonal frequency component yields the first-order temporal coherence function $\Gamma(t+t'/2,t-t'/2)$. Note that if we define a density matrix in the temporal domain $\rho^T_s(t_1,t_2)\equiv\langle t_1 |\hat{\rho}_s | t_2 \rangle$, in terms of $|t \rangle \equiv \tilde{a}^\dag(t) |0 \rangle$, it is straightforward to show that $\rho^T_s(t_1,t_2)=\Gamma^*(t_1,t_2)$, a relationship which mi\-rrors its counterpart in the spectral domain.  Inver\-ting Eq.~\ref{13} leads to a relationship between $\rho^D(\omega,\omega')$ and $W(\omega,t)$, where the roles of time and frequency are reversed with respect to those in Eq.~\ref{6}, i.e.

\begin{equation}
\label{14} W(\omega,t)=\frac{1}{4\pi^2}\int d t' \rho^T_s\left(t-\frac{t'}{2},t+\frac{t'}{2}\right) e^{-i \omega t'},
\end{equation}

\noindent which could be regarded as an alternate definition for the CWF.

%

Returning to the spectral domain, let us examine the conditions for the off-diagonal single-photon density matrix elements to be non-zero.    In terms of $\omega$ and $\omega'$, the density matrix is given by the integral over $\omega_0$ of  $f(\omega +\omega'/2,\omega_0)f^*(\omega-\omega'/2,\omega_0)$.  For a given pair of signal and idler frequencies $\omega_s=\omega$ and $\omega_i=\omega_0$, for this integrand to be non-zero, clearly the JSA function must not vanish at the symmetrically-displaced pair of signal-mode frequencies $\omega+\omega'/2$ and $\omega-\omega'/2$. Indeed, for a given $\omega$ and $\omega_0$ the width in the frequency variable $\omega'$ of the integrand is closely related to the width of the JSA function along the signal-mode frequency variable.  Note that if a one-to-one correspondence exists between the signal and idler frequencies, then this resulting width is zero, and the integrand above will be proportional to $\delta(\omega')$.  Thus, a first condition for the off-diagonal signal-mode single-photon density matrix elements to be non-zero is that each idler frequency corresponds to a \emph{spread} of signal frequencies rather than to a single signal frequency, i.e. that a strict correlation between signal and idler frequencies \textit{does not} exist.   Note that while a one-to-one correspondence in the signal and idler frequencies is consistent with maximal spectral entanglement, the correspondence of each idler frequency to a spread of signal frequencies is consistent with non-maximal spectral entanglement.

Even if each $\omega_i=\omega_0$ frequency corresponds to a spread of signal frequencies, the integration over the idler frequency $\omega_0$ implies that the single-photon density matrix may still exhibit vanishing off-diagonal elements.  Indeed, if the integrand is oscillatory in the variable $\omega_0$, integration may lead to averaging and therefore to small or va\-nishing values.  One way in which this may occur is in the case where the pump mode is chirped.    In the specific case of a quadratic chirp, quantified by parameter $\beta$, the JSA becomes

\begin{equation}
\label{15} f(\omega_s,\omega_i)=f_{0}(\omega_s,\omega_i) \mbox{exp}\left[i \beta (\omega_s+\omega_i-\omega_p^0)^2\right],
\end{equation}

\noindent where $f_{0}(\omega_s,\omega_i)$ represents the unchirped JSA.  In this case the density matrix becomes

\begin{align}
\label{16} \rho_s^D(\omega,\omega')&=e^{i 2 \beta \omega'(\omega-\omega_p^0) }\int d \omega_0 f_{0}(\omega+\frac{\omega'}{2},\omega_0) \nonumber \\
&\times f_{0}^*(\omega-\frac{\omega'}{2},\omega_0)e^{i 2 \beta \omega' \omega_0}.
\end{align}

For given non-zero $\omega'$ and $\beta$ values, the phase term in the integrand leads to oscillations as $\omega_0$ is varied, with an oscillation period given by $\pi/(\beta \omega')$.  Thus, if the integration range for $\omega_0$ is of the order, or if it exceeds this period, integration over $\omega_0$ tends to make the off-diagonal elements vanish.   A large chirp parameter $\beta$, or large $\omega'$ value (indicating matrix elements far from the diagonal) lead to a small oscillation period and thus favors small off-diagonal elements.

Thus, the existence of strong signal-idler correlations or the use of a chirped pump both lead to a small width of the density matrix $\rho_s^D(\omega,\omega')$  along $\omega'$, and therefore favor a diagonal density-matrix structure. In the case of highly  impure states,  obtained through either of these two routes, the density matrix may be expressed as

\begin{align}
\label{17} \rho^D_s \left(\omega,\omega' \right)=\frac{1}{\sqrt{\pi}\sigma'}  I_\omega(\omega) e^{-\frac{\omega'^2}{\sigma'^2}}.
\end{align}

\noindent where $\sigma'$ is a small quantitity.  For a strictly impure state with $\sigma' \rightarrow 0$ the CWF is then, according to the Fourier transform relationship of Eq.~\ref{6}, time independent i.e.

\begin{align}\label{18}
W(\omega,t) \propto I_\omega(\omega),
\end{align}

Thus, a perfectly impure state leads to a constant temporal intensity profile, or in other words to an infinite single-photon temporal duration. Also, note from Eq.~\ref{13} and Eq.~\ref{18} that such a perfectly impure state leads to a first degree temporal coherence function of the following form

\begin{equation}
\label{19} \Gamma(t_1,t_2) \propto  \int d \omega I_\omega(\omega) e^{i \omega(t_1-t_2)}.
\end{equation}

Note that the above $\Gamma(t_1,t_2)$ depends on the two arguments only through their difference. In other words, a perfectly impure single-photon state is statistically temporally stationary.  In general terms, there is a link between the temporal duration of the signal photon and its degree of purity. In order to study this link, let us consider the temporal intensity profile $I_t(t)$ of the signal-mode single photon.  This function is given by

\begin{align}\label{20}
I_t(t) &\equiv \int d \omega W(\omega,t) \nonumber \\
&= \frac{1}{2\pi} \int d \omega' \int d \omega \rho_s^D \left(\omega,\omega'\right) e^{-i \omega't }.
\end{align}

Normalization of $I_t(t)$, i.e. $\int dt I_t(t)=1$ is guaranteed by the normalization of the density matrix, i.e. $\mbox{Tr}(\hat{\rho}_s)=1$.   As is clear from Eq.~\ref{20}, this temporal profile is given by the Fourier transform of the density matrix averaged over all $\omega$ values.   For an increasingly impure single-photon state, leading to a density matrix with an increasingly diagonal structure, the width of the $\int d \omega \rho_s^D \left(\omega,\omega'\right)$ function along $\omega'$ is reduced, leading to an increased  signal-mode temporal duration as dictated by the Fourier transform relationship of Eq.~\ref{20}.   This is consistent with the analysis in Ref.~\cite{uren07} carried out in the context of the Gaussian approximation for the JSA, leading to the result $\delta t \delta \omega= 1/p$, where $\delta t$ and $\delta \omega$ are the temporal duration and spectral width of the signal-mode single photon, respectively, and $p$ is the single-photon purity.  Thus, for example, as the pump chirp $\beta$ is increased, while $\delta \omega$ (the width of the marginal distribution of the joint spectrum $|f(\omega_s,\omega_i)|^2$) remains fixed, since the two-photon state depends on $\beta$ only through a phase, the temporal duration $\delta t$ is increased, leading to a an increased product $\delta t \delta \omega$, and consequently to a reduced purity $p$.

The relationship between photon pair-entanglement and single-photon purity is well understood. In fact, single photon purity is simply the reciprocal of the Schmidt number $K$, which quantifies the degree of photon-pair entanglement. This underscores the importance of factorable states, with $K=1$, which lead to pure states, with $p=1$.  Conversely, highly-entangled photon pairs, with $K \rightarrow \infty$, lead to highly impure single photons with $p \rightarrow 0$~\cite{uren05}.

The joint spectral amplitude $f(\omega_s,\omega_i)$ can be written as follows

\begin{equation}
\label{21} f(\omega_s,\omega_i)=\phi(\omega_s,\omega_i) \alpha(\omega_s+\omega_i)  e^{i \beta (\omega_s+\omega_i-\omega_p^0)^2},
\end{equation}

\noindent in terms of the phasematching function (PMF)  $\phi(\omega_s,\omega_i)$
which describes the optical properties of the nonlinear crystal, and the pump
envelope function (PEF) $\alpha(\omega)e^{i \beta (\omega-\omega_p^0)^2}$ which
explicitly includes a quadratic chirp phase.  The PMF may be written as
$\phi(\omega_s,\omega_i)=\mbox{sinc}[L \Delta k(\omega_s,\omega_i)/2]$, in terms of the crystal length $L$ and of the phasemismatch $\Delta k(\omega_s,\omega_i)=k_p(\omega_s+\omega_i)-k_s(\omega_s)-k_i(\omega_i)$; here $k_\nu$ (with $\nu=p,s,i$ represents the wavenumber for each of the pump(p), signal(s), and idler(i).

In order to facilitate the analysis below, it is helpful to rely on an analytical expression for the CWF.  As discussed in Ref.~\cite{uren07}, this can be achieved through two approximations: i) expressing the JSA function entirely in terms of Gaussian functions, and ii) relying on a power series expansion, truncated at first order, of the Gaussian-approximated JSA argument. In this case, we can write the JSA function expressed in terms of detu\-nings $\nu_{s,i}=\omega_{s,i}- \omega_{s,i}^0$ (where $\omega_{s,i}^0$ represent signal/idler central phasematched frequencies) as

\begin{eqnarray}\label{22}
f(\nu_s, \nu_i) &=& e^{-[X^2_{ss}\nu_s^2 + X^2_{ii} \nu_i^2 + 2(X_{si}^2-i \beta)\nu_s\nu_i]}.
\end{eqnarray}

In Eq.~\ref{22} we have used the following definition,
with $\lambda,\mu=s,i$: $X_{\lambda \mu}\equiv\frac{1}{\sigma^2}
+ \frac{\gamma\tau_\lambda\tau_\mu}{4}$, in turn written
in terms of the pump bandwidth $\sigma$, the pump chirp parameter $\beta$,
and parameter $\gamma=0.193$ (related to the Gaussian
approximation of the JSA, see Ref.~\cite{uren07}). This definition
is also in terms of the signal/idler group velocity mismatch
parameters $\tau_\lambda$ (with $\lambda=s,i$) given in terms
of the crystal length $L$ as $\tau_\lambda=L(k_p'-k_\lambda')$;
here, $k'$ denotes the frequency derivative of the wavenumber
associated to the each of the modes ($p,s,i$), evaluated at the
 respective central frequency.

Carrying out the corresponding integrations (Eqns.~\ref{3} and \ref{6}),
we obtain the following expression for the CWF valid, for $\tau_s \neq \tau_i$

\begin{eqnarray}\label{23}
W_s(\nu, t) &=& \frac{\sqrt{1-\mathscr{C}^2}}{\pi\Delta t \Delta \omega}e^{-\left(\frac{\nu}{\Delta
\omega}\right)^2}e^{-\left(\frac{t}{\Delta
t}\right)^2} e^{-2 \mathscr{C} \left(\frac{\nu}{\Delta
\omega}\right) \left (\frac{t}{\Delta
t}\right)}.
\end{eqnarray}

The parameters in Eq.~\ref{23} include a spectral width parameter $\Delta \omega$, its square given by

\begin{eqnarray}\label{24}
\label{24} \nonumber \Delta \omega^2=\frac{\beta^2+X_{ss}X_{ii}}{2\left[\beta^2(X_{ss}-2 X_{si}+X_{ii}) +
X_{ss}(X_{ss}X_{ii}-X_{si}^2)\right]},\\
\end{eqnarray}

\noindent a temporal width parameter $\Delta t$, its square given by

\begin{equation}\label{25}
\Delta t^2=\frac{2(\beta^2+X_{ss}X_{ii})}{X_{ii}},
\end{equation}

\noindent and a single-photon chirp parameter $\mathscr{C}$ given by

\begin{equation}\label{26}
 \mathscr{C}=\frac{\beta(X_{ii}-X_{si})}{\sqrt{X_{ii}[\beta^2(X_{ss}-2 X_{si}+X_{ii})
+X_{ss}(X_{ss} X_{ii}-X_{si}^2)]}}
\end{equation}

When plotted in a normalized chronocyclic space $\{\nu/\Delta \omega,
t/\Delta t\}$, the CWF has a circular shape for $\mathscr{C}=0$ and becomes
elongated diagonally for $\mathscr{C} \neq 0$. In par\-ticu\-lar, for
$0<\mathscr{C}<1$ the distribution involves correlated frequencies and times,
while for $-1<\mathscr{C}<0$ the distribution involves anti-correlated
frequencies and times. For $|\mathscr{C}|=1$, the distribution becomes infinitely elongated,
indicating the maximum degree of single-photon chirp. Using the inequality
$X_{si}^2 \le X_{ss} X_{ii}$, which may be readily proved, it is
straightforward to show that Eq.~\ref{26} indeed fulfils $|\mathscr{C}| \le 1$.

In the limiting case where $\tau_s \rightarrow \tau_i$, corresponding to frequency
anti-correlated photon pairs, $X_{si} \rightarrow X_{ss}=X_{ii}$ and it may be shown that $\Delta \omega \rightarrow \infty$
and $\mathscr{C} \rightarrow 0$. This resulting infinite spectral width appears because
for anti-correlated photon pairs the phasemathcing function and the pump envelope function have the same
orientation in $\{\omega_s,\omega_i\}$ space, and within the linear approximation of the phasemismatch used for the analytic
expression of the CWF (Eqns.~\ref{23} to \ref{26}), the curvature of the phasematching function is suppressed.
In a realistic i.e. unapproximated situation, the non-zero curvature leads to a finite (but large) spectral width.

As is evident from Eq.~\ref{26}, in general in the absence of pump chirp, the
signal-mode single photons are likewise unchirped.  The presence of pump chirp can lead to
single-photon chirp, depending on the type of spectral correlations in the SPDC
photon pairs.  In order to make this more evident, we can write
the numerator of Eq.~\ref{26} proportional to
$\beta \tau_i(\tau_s-\tau_i)$. As discussed in Ref.~\cite{uren05}, a factorable state with
an elongated joint spectrum is possible if $\tau_s=0$ or $\tau_i=0$. Thus, interestingly,
a factorable, spectrally elongated state with $\tau_i=0$ is such that the signal-mode
single photons remain unchirped despite an arbitrary level of pump chirp.  Likewise a state with spectral
anti-correlations, characterized by $\tau_s=\tau_i$ which as discussed in the previous paragraph
leads to $\mathscr{C} \rightarrow 0$, is also such that that the signal-mode single photons
remain unchirped despite an arbitrary level of pump chirp (in this case, Eq.~\ref{26} should be evaluated in the limit where $X_{si} \rightarrow X_{ss}=X_{ii}$). In general, for other types of
joint spectra, the single photons become increasingly chirped as the level of pump
chirp is increased.

We are interested in studying how certain key photon-pair properties determine the resulting single-photon properties. In particular, we will consider the follo\-wing two photon-pair properties: i) type of spectral correlations present as determined by the joint spectrum $|f(\omega_s,\omega_i)|^2$, and ii)level of quadratic pump chirp present, quantified through the parameter $\beta$. In turn, we are particularly interested in studying through the single-photon density matrix and the CWF, the single-photon purity and the type of resulting chirp in the single photons. We will consider five representative types of two-photon state, as characterized by the type of spectral correlations present: i) spectral anti-correlations, ii) spectral positive correlations, iii) factorable state with a vertically-oriented joint spectrum, iv) factorable state with a horizontally-oriented joint spectrum and iv) factorable state with a symmetric joint spectrum.

\begin{figure}[h!]
\begin{center}
\centering\includegraphics[width=9cm]{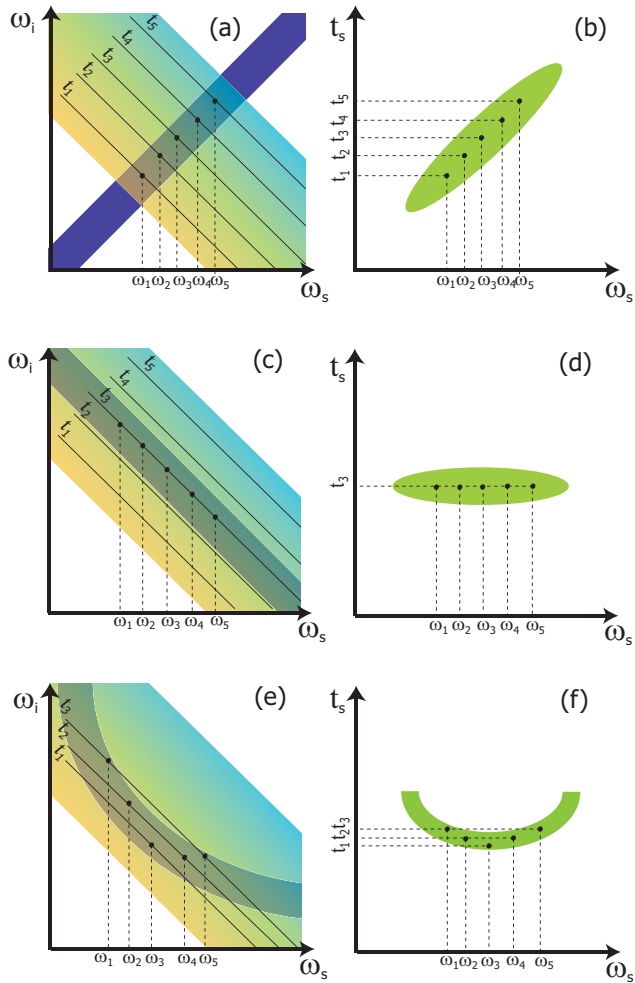}
\end{center}
\par
\caption{Panels a, c, and e show
schematically  how the joint spectrum $|f(\omega_s,\omega_i)|^2$ is determined by
the phasematching function $|\phi(\omega_s,\omega_i)|^2$ and the pump envelope
function $|\alpha(\omega_s+\omega_i)|^2$.  These three panels correspond to: a) positive correlations, c) anti-correlations with curvature suppressed, e)anti-correlations.  Panels b), d) and f) show the corresponding inferred single-photon chronocylic structure.
} \label{caricatura1}
\end{figure}

\begin{figure}[h!]
\begin{center}
\centering\includegraphics[width=9cm]{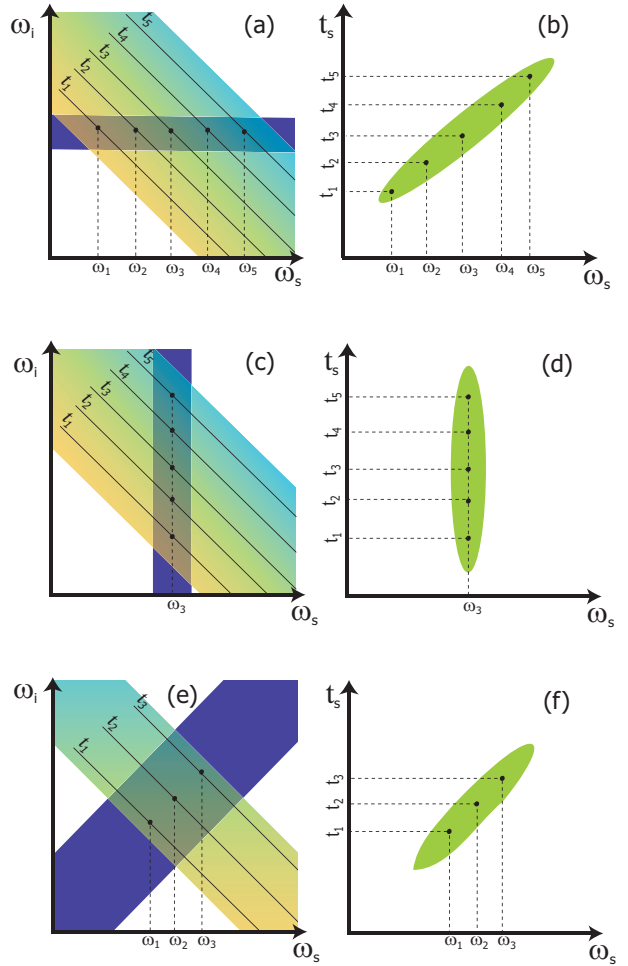}
\end{center}
\par
\caption{Panels a, c, and e show
schematically  how the joint spectrum $|f(\omega_s,\omega_i)|^2$ is determined by
the phasematching function $|\phi(\omega_s,\omega_i)|^2$ and the pump envelope
function $|\alpha(\omega_s+\omega_i)|^2$.  These three panels correspond to: a) factorable state with horizontally-oriented joint spectrum, c) factorable state with vertically-oriented joint spectrum, and e) factorable state with circular-shaped joint spectrum.  Panels b), d) and f) show the corresponding inferred single-photon chronocylic structure. } \label{caricatura2}
\end{figure}

Let us consider each of these types of two-photon states, in turn. Figs.~\ref{caricatura1} and \ref{caricatura2} show schematically for these states how the joint spectrum is determined by the functions $\phi(\omega_s,\omega_i)$ and $\alpha(\omega_s+\omega_i)$. These figures also show schematically how quadratic pump chirp, together with the unchirped joint spectrum determine the chronocyclic structure of the single photons which constitute each of the photon pairs.  In Fig.~\ref{caricatura1} each of the three rows co\-rres\-pond to the following states: positively-correlated state, anti-correlated state, and a version of the anti-correlated state exhibiting some curvature in the joint spectrum.  In each row, the left-hand panel corresponds to a re\-pre\-sen\-tation of the joint spectrum, with the blue band representing the phasematching function $\phi(\omega_s,\omega_i)$  and the gradient-colored band indicating the pump envelope function $\alpha(\omega_s,\omega_i)$.  Note that the PEF exhibits the same orientation in $\{\omega_s, \omega_i\}$ space for all three states, and in all cases the width indicates the use of a broadband, pulsed pump. However, the PMF shows a different orientation in each of the cases, determined by the specific crystal configuration used, which in part determines the resulting spectral entanglement characteristics.

The color gradient used to shade the PEF indicates the presence of quadratic pump chirp, in which case different pump frequencies correspond to different temporal ``slices'' of the temporally-broadened pump pulse. Thus,  we can think of the chirped pump as composed of a temporal sequence of essentially single-frequency components. Each of these single-frequency components, $\Omega$, corresponds to the locus $\omega_s+\omega_i=\Omega$ in $\{\omega_s, \omega_i\}$ space.  In this manner, as time progresses within a given pulse, the pump probes different diagonal slices of the joint spectrum, i.e. the joint spectrum produced at a given pump pulse slice with frequency $\Omega$ can be written as $|\phi(\omega_s,\omega_i)\alpha(\omega_s+\omega_i)|^2\delta(\omega_s+\omega_i-\Omega)$.  The presence of chirp means that each $\Omega$ occurs at a different time, indicated schematically in Figs.~\ref{caricatura1} and \ref{caricatura2} by black diagonal lines.

From this information, it is possible to infer the chronocyclic structure of, say, the signal-mode single photon.  As time progresses within a given quadratically-chirped pump pulse, a different diagonal slice of the joint spectrum is emitted.  The signal-mode spectrum emitted at a given time is given by the intersection of the locus $\omega_s+\omega_i=\Omega$ (where $\Omega$ is time-dependent) with the unchirped joint spectrum $|\phi(\omega_s,\omega_i) \alpha(\omega_s+\omega_i)|^2$. The right panel on each row of Figs.~\ref{caricatura1} and \ref{caricatura2} shows the inferred choronocyclic structure of the signal-mode single photon, obtained by plotting the points shown within each joint spectrum (left panels of these figures) in the signal-mode chronocyclic space $\{\omega_s,t_s\}$.

There is clear relationship between the type of photon-pair
 spectral correlations and the resulting single-photon chronocyclic properties. Let us review each of the types of spectral correlations. For photon pairs with positive correlations, there is a monotonic resulting relationship between
 the time within the pump pulse and the resulting signal-mode frequency. This results in temporal-spectral correlations in the signal-mode single photon, or in other words the signal-mode single photon becomes chirped. These correlations may be positive for positive $\beta$ or ne\-ga\-tive for negative $\beta$. In contrast, for vanishing pump chirp the signal-mode single photon is likewise unchirped showing no spectral-temporal correlations.

Let us now turn our attention to spectrally anti-correlated photon pairs.
In the idealized case of strict spectral anti-correlations, there is a
single pump pulse time, related to a single pump frequency $\Omega$ that
participates in the SPDC process. Thus, in this case despite the
presence of pump chirp, all signal-mode frequencies co\-rres\-pond to a single time of emission, and thus arbitrarily strong pump chirp does not translate into single-photon chirp. Note that this is consistent with the conclusions
reached in terms of our analytic experssion for the CWF (see Eq.~\ref{23});
in this case, $\tau_s=\tau_i$ means that $\mathscr{C} \rightarrow 0$ despite the presence of
pump chirp. In a realistic situation for a broadband pump,
the joint spectrum exhibits some curvature related to group velocity
dispersion as shown schematically in Fig.~\ref{caricatura1}(e). In this case, the locus $\omega_s+\omega_i-\Omega=0$ will
intersect the joint spectrum at two distinct areas which approach each other,
and even\-tua\-lly merge into a single zone, as $\Omega$ is reduced. The \-effect\- of this is that the curvature which characterizes the joint spectrum translates into a curvature in the single-photon chronocyclic structure, as shown schematically in Fig.~\ref{caricatura1}(f).

Let us now turn our attention to factorable photon pairs;
we will consider three different kinds of factorable states:
those with a horizontally-oriented joint spectrum, those with
a vertically-oriented joint spectrum, and those with a symmetric
joint spectrum.  As may be seen from Fig.~\ref{caricatura2}(a), in the case
of a  horizontally-oriented joint spectrum, the diagonal lines which
indicate different pump pulse times intersect the joint
spectrum at distinct points. Thus, in this case in a
manner qualitatively similar to the positive-correlation case,
there is a resulting correlation between the emission time and the
signal-mode emission frequency. In the case of a vertically-oriented joint
spectrum, the situation is different. Here, see Fig.~\ref{caricatura2}(b),
the different emission times of the selected points all correspond to the same signal-mode
emission frequency. This implies that the presence of arbitrarily
large pump chirp does not result in signal-mode single-photon chirp.
Note that this is consistent with the conclusions
reached in terms of our analytic expression for the CWF (see Eq.~\ref{23});
in this case, $\tau_i=0$ means that $\mathscr{C}=0$ despite the presence of
pump chirp.

In what follows, we present specific numerical calculations showing
the density matrix on the one hand and the chronocyclic Wigner function
on the other hand for each type of photon-pair spectral correlations, and
also showing the effect of quadratic pump chirp.  Note that for these
numerical simulations we use the full, unapproximated joint amplitude function.
As we will show, the intuition gained both from the joint spectrum
schematics (Figs.~\ref{caricatura1} and \ref{caricatura2}) and from the analytic expression
of the CWF (see Eq.~\ref{23}) agree well with these numerical results.   We will consider a
specific implementation for each of the types of source considered above.
Table I shows for each these sources: the type of crystal used, the configuration used (type I or type II),
the crystal length $L$, the crystal cut angle $\theta_{pm}$,  the crystal phasematching bandwidth $\Delta \omega_c$,  the pump center
wavelength, $\lambda_o$,  the pump bandwidth $\Delta \omega$ and the Fourier-transform limited pulse duration $\tau$.  Note that we have used as definition of
 $\Delta \omega_c$ the full width at half maximum of the function $|\phi(\omega/2,\omega/2)|^2$,  which may  also be thought of as the pump-frequency acceptance function of the crystal.

Photon pair properties are determined by the characteristics of the crystal and
of the pump.  In particular, if
$\Delta \omega_c< \Delta \omega$,  the phase matching function dominates
over the pump envelope function to determine the two-photon state, and if
 $\Delta \omega< \Delta \omega_c$, the pump envelope function dominates over
 the phase matching function. As may be seen in the table, among the particular states chosen for illustration purposes, for those  characterized by an anti-correlated, horizontal
 and vertical joint spectrum, $\Delta \omega_c< \Delta \omega$, whereas for the positive-correlation and
 the factorable with a circular joint spectrum states, $\Delta \omega_c>\Delta \omega$.

In all sources considered here, see  Table I, we assume frequency-degenerate, collinear SPDC.  Likewise,
in all cases involving non-zero pump chirp, we have assumed a value
of $\beta=8 \times10^{-26}$s$^2$.

\begin{widetext}
\begin{center}
\begin{table}[ht]\centering\label{table}
 \caption{PARAMETERS}
    \begin{tabular}{|c| c  c c c |c c c|}
    \hline
 Correlation &  & Crystal & characteristics & &  & Pump & \\ \cline{2-8}
 & Type & L & $\theta_{PM}$ & $\Delta\omega_c$ & $\lambda_o$ & $\Delta\lambda$ ($\Delta\omega$) & $\tau$\\
\hline\hline
\centering Horizontal & \centering KDP-II & \centering 5 mm  & \centering $67.8^o$ & \centering 15.6 THz & \centering 415 nm & 5 nm (54.7 THz) & 50.7 fs \\
\hline
\centering Vertical & \centering KDP-II &\centering 5 mm & \centering $67.8^o$ & \centering 15.6 THz &
\centering 415 nm & 5 nm (54.7 THz) &  50.7 fs\\
\hline
\centering Positive & \centering BBO-II  & \centering 10 mm & \centering $28.8^o$ & \centering 208.2 THz & \centering 757 nm & 20 nm (65.8 THz) & 42.1 fs\\
\hline
\centering Negative  & \centering BBO-I & \centering 2 mm &\centering $29.2^o$  & \centering 28.3 THz&
\centering 400 nm &  5 nm (58.9 THz) & 47.1 fs\\
\hline
\centering Circular  & \centering BBO-II & \centering $2.293$ mm & \centering $28.8^o$ & \centering 328.9 THz & \centering 757 nm & 15 mm (49.3 THz) & 56.2 fs \\
\hline
\bottomrule
    \end{tabular}
\end{table}
\end{center}
\end{widetext}

\begin{figure}[h!]
\begin{center}
\centering\includegraphics[width=7cm]{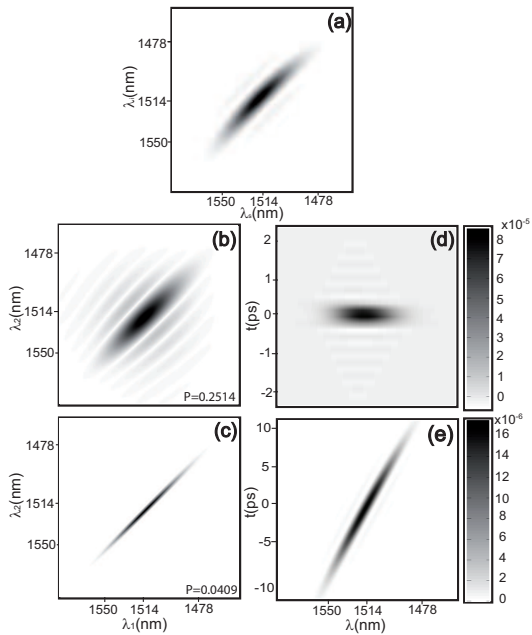}
\end{center}
\par
\caption{For a state with positive spectral correlations: a) joint spectrum $|f(\omega_s,\omega_i)|^2$, b) signal-mode, single-photon density matrix $|\rho_s(\omega_1,\omega_2)|$ with the single-photon purity value indicated, c) same as b) but with pump chirp $\beta=8 \times 10^{-26} s^2$, d) single-photon chronocyclic Wigner function $W(\omega,t)$, e) same as d) but with pump chirp $\beta=8 \times 10^{-26} s^2$. Note: for convenience frequency axes are labeled in terms of wavelength.}\label{positive}
\end{figure}

Fig.~\ref{positive} corresponds to the state with positive correlations;  panel (a) shows the joint spectrum.  The entanglement present implies that the signal-mode single photons are impure, as indicated by the diagonal structure of the density matrix; panel (b) is a plot of $|\rho_s(\omega_1,\omega_2)|$, for $\beta=0$.   The presence of pump chirp further reduces the width along the anti-diagonal, i.e. $\omega'$, direction of the density matrix, as discussed above due to averaging, see panel (c).  In the absence of pump chirp, signal-mode single photons are unchirped, i.e. emission times and frequencies are uncorrelated, as is clear from the CWF plotted in panel (d).  The presence of pump chirp, as discussed above for this state, has the effect of chirping the signal-mode single photon as is clear from the time-frequency correlations in the CWF plotted in the presence of chirp, see panel (e).

\begin{figure}[h!]
\begin{center}
\centering\includegraphics[width=7cm]{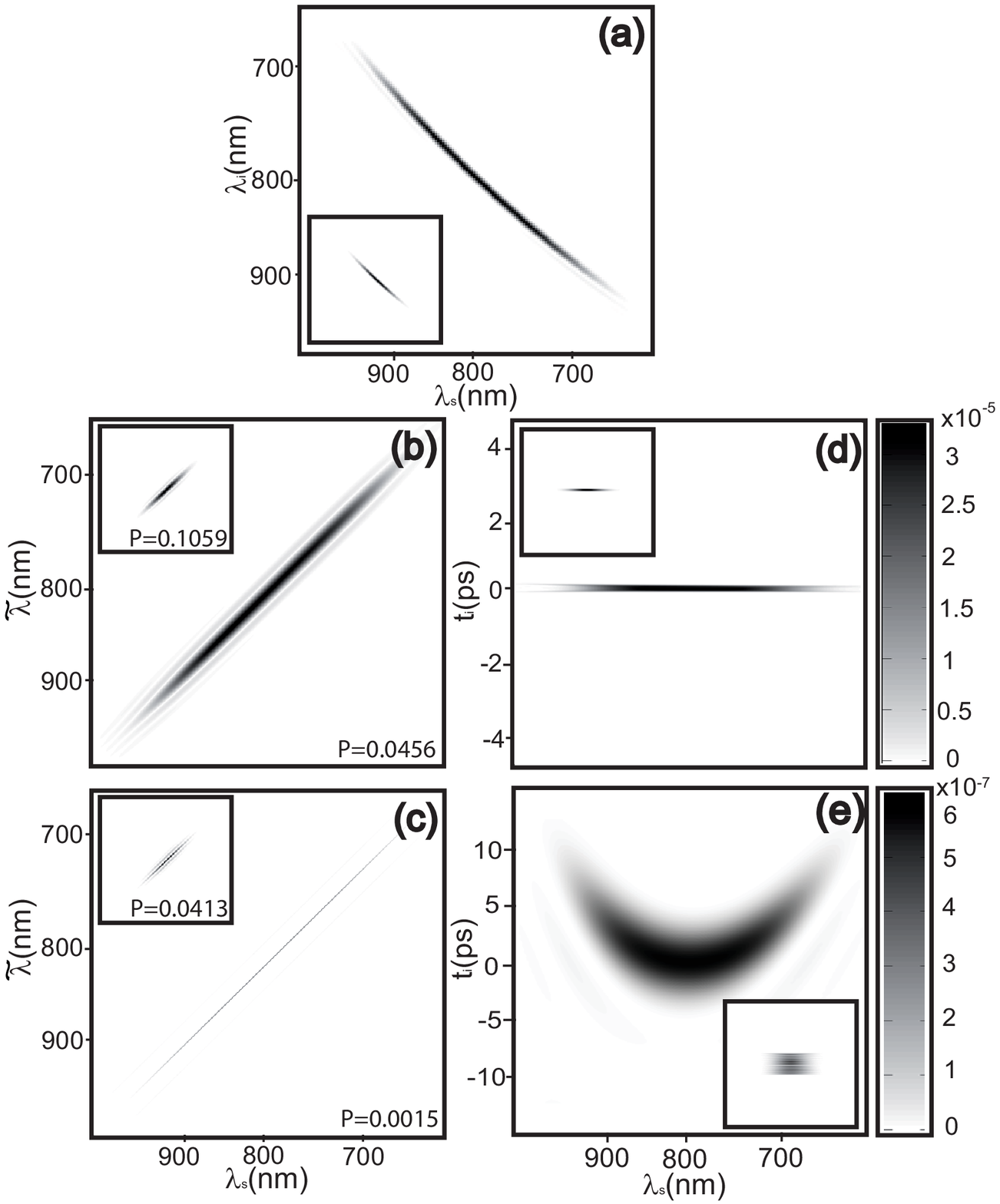}
\end{center}
\par
\caption{For a state with negative spectral correlations: a) joint spectrum $|f(\omega_s,\omega_i)|^2$, b) signal-mode, single-photon density matrix $|\rho_s(\omega_1,\omega_2)|$ with the single-photon purity value indicated, c) same as b) but with pump chirp $\beta=8 \times 10^{-26} s^2$, d) single-photon chronocyclic Wigner function $W(\omega,t)$, e) same as d) but with pump chirp $\beta=8 \times 10^{-26} s^2$.  Insets show the result of transmitting the signal and idler photons through a Gaussian-profile spectral filter with $100$nm bandwidth. Note: for convenience frequency axes are labelled in terms of wavelength.} \label{negative}
\end{figure}

Fig.~\ref{negative} corresponds to the spectrally anti-correlated
state; panel (a) shows the joint spectrum.  Note that the joint spectrum
exhibits some curvature in the $\{\omega_s, \omega_i\}$ space, related to
group velocity dispersion of the SPDC photon pair. The emission bandwidth,
i.e. the width of the SPS, of $0.23\mu$m is considerable. The inset of panel
(a) shows the joint spectrum assuming that the signal and idler photons
are each transmitted through a Gaussian-profile bandpass filter with $100$ nm
bandwidth; as can be appreciated, the curvature is then essentially suppressed.
As in the case of positive correlations, the entanglement present implies
that the signal-mode single photons are impure as indicated by the diagonal
structure of the density matrix; panel (b) is a plot of
$|\rho_s(\omega_1,\omega_2)|$, for $\beta=0$.  Again, the presence of pump
chirp further reduces the width the along the anti-diagonal of the density
matrix due to averaging, see panel (c).  As for the positively-correlated
case, in the absence of chirp, the signal-mode single photons are unchirped,
i.e. emission times and frequencies are uncorrelated, as is clear from the
CWF plotted in panel (d).  As discussed above, the anti-correlated state
is relatively insensitive to pump chirp.   Let us first consider the
filtered state shown in the inset of panel (a).  The effect of pump
chirp on the CWF for this filtered, anti-correlated state, may be seen
in the inset of panel (e); it is clear that despite the presence of pump
chirp, there is no resulting chirp in the signal-mode single photons.
In the case where spectral filtering is not used, the CWF shows a curvature which,
as discussed above, is related to the curvature of the joint spectrum; see
panel (e).

\begin{figure}[h!]
\begin{center}
\centering\includegraphics[width=7cm]{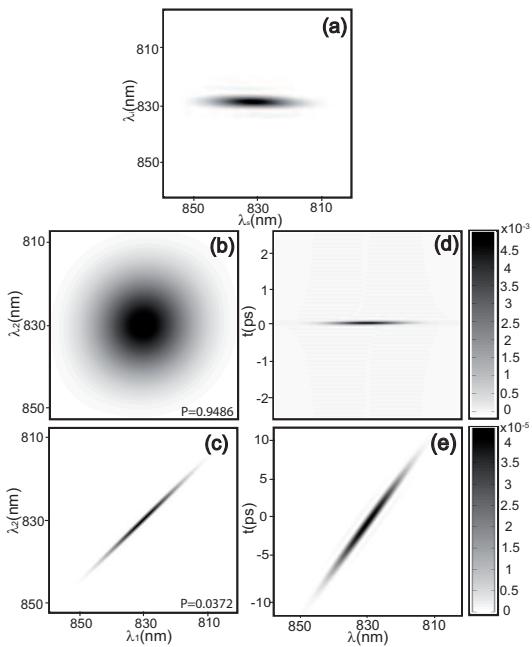}
\end{center}
\par
\caption{For a factorable state with a horizontally-oriented joint spectrum: a) joint spectrum $|f(\omega_s,\omega_i)|^2$, b) signal-mode, single-photon density matrix $|\rho_s(\omega_1,\omega_2)|$ with the single-photon purity value indicated, c) same as b) but with pump chirp $\beta=8 \times 10^{-26} s^2$, d) single-photon chronocyclic Wigner function $W(\omega,t)$, e) same as d) but with pump chirp $\beta=8 \times 10^{-26} s^2$. Note: for convenience frequency axes are labelled in terms of wavelength.} \label{horizontal}
\end{figure}

\begin{figure}[h!]
\begin{center}
\centering\includegraphics[width=7cm]{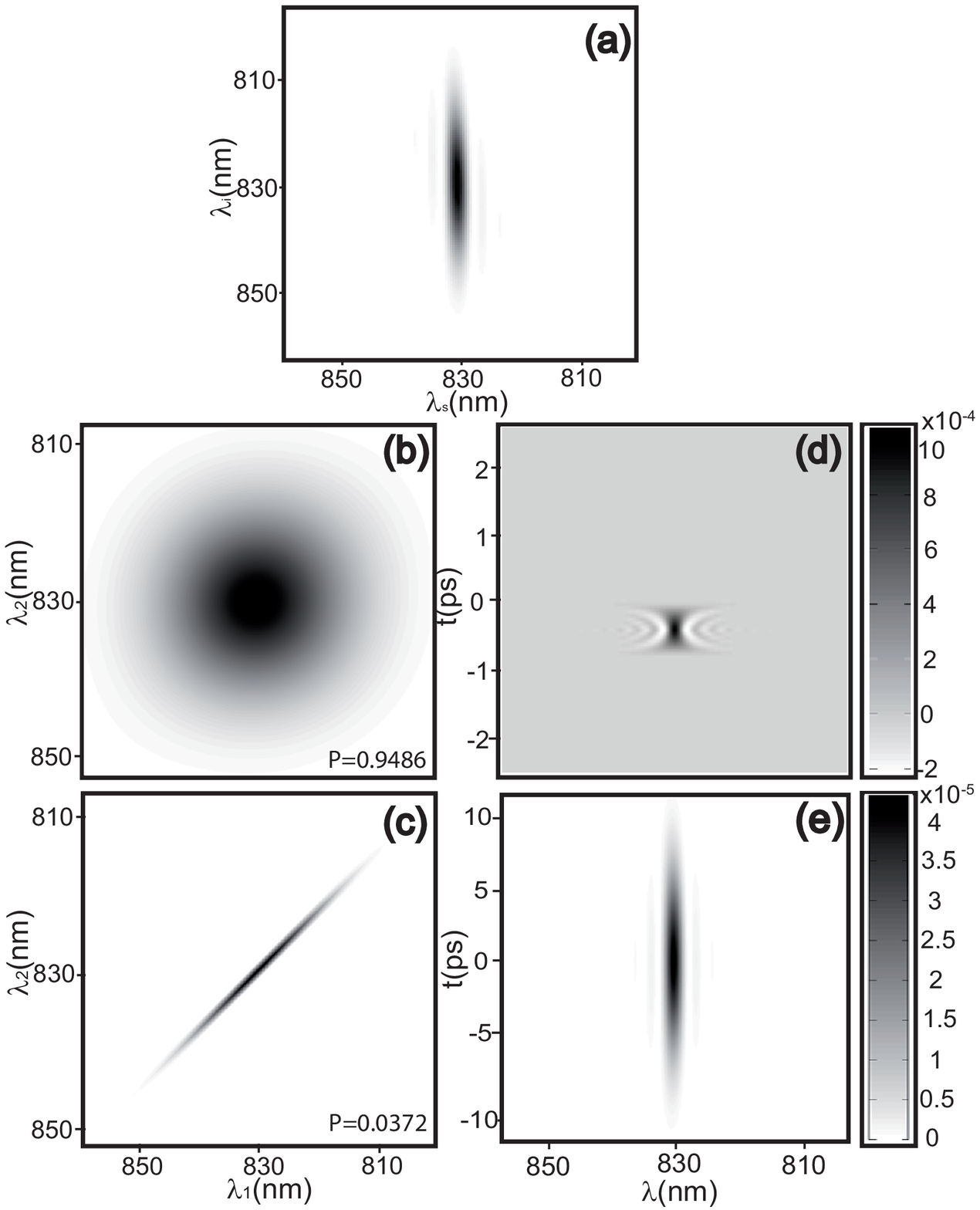}
\end{center}
\par
\caption{For a factorable state with a vertically-oriented joint spectrum: a) joint spectrum $|f(\omega_s,\omega_i)|^2$, b) signal-mode, single-photon density matrix $|\rho_s(\omega_1,\omega_2)|$ with the single-photon purity value indicated, c) same as b) but with pump chirp $\beta=8 \times 10^{-26} s^2$, d) single-photon chronocyclic Wigner function $W(\omega,t)$, e) same as d) but with pump chirp $\beta=8 \times 10^{-26} s^2$. Note: for convenience frequency axes are labelled in terms of wavelength.} \label{vertical}
\end{figure}

\begin{figure}[h!]
\begin{center}
\centering\includegraphics[width=7cm]{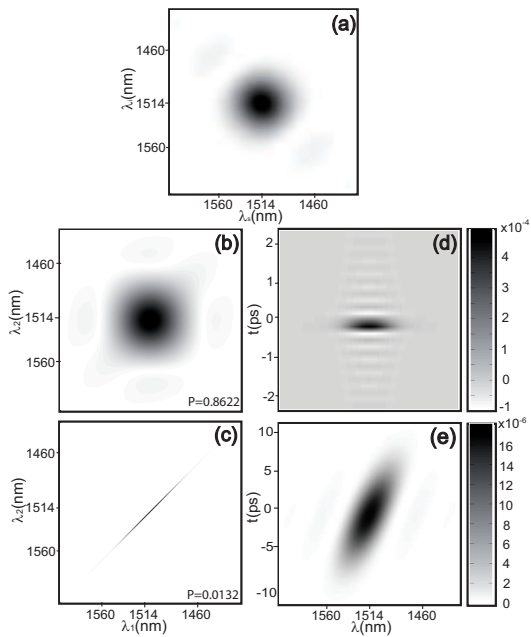}
\end{center}
\par
\caption{For a factorable state with a circular-shaped joint spectrum: a) joint spectrum $|f(\omega_s,\omega_i)|^2$, b) signal-mode, single-photon density matrix $|\rho_s(\omega_1,\omega_2)|$ with the single-photon purity value indicated, c) same as b) but with pump chirp $\beta=8 \times 10^{-26} s^2$, d) single-photon chronocyclic Wigner function $W(\omega,t)$, e) same as d) but with pump chirp $\beta=8 \times 10^{-26} s^2$. Note: for convenience frequency axes are labelled in terms of wavelength.} \label{circular}
\end{figure}

Fig.~\ref{horizontal} corresponds to the factorable state, with a horizontally-oriented joint spectrum; panel (a) shows a plot of the joint spectrum.  The nearly-factorable character of the photon-pair state implies that the signal-mode single photons are basically pure in the absence of pump chirp.  This in turn implies that some of the coherences, i.e. the off-diagonal elements of the density matrix are non-zero.  This is evident in panel (b), which represents a plot of $|\rho_s(\omega_1,\omega_2)|$.  The presence of pump chirp has the expected effect of averaging out the off-diagonal elements, so that the density matrix acquires a diagonal structure, as shown un panel (c).  In the absence of pump chirp, signal-mode single photons are unchirped, i.e. emission times and frequencies are uncorrelated, as is clear from the CWF, plotted in panel (d).  The presence of pump chirp, as discussed above for this state, has the effect of chirping the signal-mode single photon as is clear from the time-frequency correlations in the CWF plotted for $\beta \neq 0$, see panel (e).

Fig.~\ref{vertical} corresponds to the factorable state, with a vertically-oriented joint spectrum; panel (a) shows a plot of the joint spectrum.  The nearly-factorable character of the photon-pair state implies that the signal-mode single photons are basically pure in the absence of pump chirp.  As for the factorable state with a horizontally-oriented joint spectrum, this in turn implies that some of the coherences, i.e. the off-diagonal elements of the density matrix are non-zero.  This is evident in panel (b), which represents a plot of $|\rho_s(\omega_1,\omega_2)|$. The presence of pump chirp has the expected effect of ave\-ra\-ging out the off-diagonal elements, so that the density matrix acquires a diagonal structure, as shown un panel (c).  In the absence of pump chirp, signal-mode single photons are unchirped, i.e. emission times and frequencies are uncorrelated, as is clear from the CWF, plotted in panel (d).  As discussed above, this state has the property that despite the presence of pump chirp, the signal-mode single photons remain basically unchirped, as is evident from the CWF with $\beta \neq 0$, plotted un panel (e).

Finally, Fig.~\ref{circular} corresponds to the factorable state, with a symmetric, or circularly-shaped, joint spectrum;  panel (a) shows a plot of the joint spectrum.  The nearly-factorable character of the photon-pair state implies that the signal-mode single photons are basically pure in the absence of pump chirp.  As for the previous two types of factorable states above, this in turn implies that for $\beta=0$ some of the coherences, i.e. the off-diagonal elements of the density matrix are non-zero.  This is evident in panel (b), which represents a plot of $|\rho_s(\omega_1,\omega_2)|$.  The presence of pump chirp has the expected effect of averaging out the off-diagonal elements, so that the density matrix acquires a diagonal structure, as shown un panel (c).  In the absence of pump chirp, signal-mode single photons are unchirped, i.e. emission times and frequencies are uncorrelated, as is clear from the CWF, plotted in panel (d).  In the presence of pump chirp, the signal-mode single photons the signal-mode single photons acquire a chirp as is evident from the CWF with $\beta \neq 0$, plotted un panel (e).

\section{CONCLUSIONS}

In this paper we have studied the chronocyclic properties of the single-photon constituents of photon pairs ge\-nera\-ted by the process of spontaneous parametric down conversion. We have studied how photon-pair properties, in particular, i)the type of signal-idler spectral correlations, and ii) pump chirp, determine the resulting single-photon properties, in particular the purity and the single-photon chirp. We have studied single-photon properties through the corresponding single-photon spectral density matrix $\rho_s(\omega_1,\omega_2)$, and in chronocyclic space through the single-photon chronocyclic Wigner function $W(\omega,t)$, and how these are determined by photon pair properties, in\-clu\-ding the joint spectrum $|f(\omega_s,\omega_i)|^2$ and the level of pump chirp $\beta$. We have studied the relationship between these functions and the first degree of spectral coherence $S(\omega_1,\omega_2)$ on the one hand, and the first degree of temporal cohe\-rence $\Gamma(t_1,t_2)$ on the other hand.

As we have shown before, pump chirp may be used as an effective tool to control photon-pair entanglement, or equivalently to control the single-photon purity. Varying the level of pump chirp in order to control spectral entanglement, also results in a modified chronocyclic single-photon character. In this paper we focus on the study of this single-photon chronocyclic character, in particular the single-photon chirp, and how this is determined by photon pair properties. We find that besides the previously-understood link between pump chirp and degree of photon pair entanglement or single photon purity, pump chirp in general leads to single-photon chirp, in a manner determined by the type of spectral correlations present. In fact, for certain types of signal-idler spectral correlations single photons remain unchirped despite the presence of pump chirp. We hope that this paper will contribute to a better appreciation of the multi-mode, spectral and temporal, character of single photons derived from spontaneous parametric downconversion.

\begin{acknowledgements}
This work was supported in part by CONACYT, Mexico,  by DGAPA, UNAM
and by FONCICYT project 94142.  We would like to thank Y. Jeronimo-Moreno and K. Garay-Palmett, for stimulating discussions.
\end{acknowledgements}

\end{document}